\documentclass[aps,prx,10pt,amssymb,amsmath,superscriptaddress,tightenlines,twocolumn,notitlepage,
] {revtex4-2}
\usepackage{graphicx}
\usepackage{amsmath}
\usepackage{amssymb}
\usepackage{bm}
\usepackage{color}
\usepackage{hyperref}
\usepackage{tabularx}
\usepackage{multirow}
\usepackage{booktabs}
\usepackage{braket}
\usepackage{epstopdf}
\newcommand{\SZL}[1] {{\color{red}#1}}

\begin{document}

\title{Chiral superconductivity from spin polarized Chern band in twisted MoTe$_2$}

\author{Cheng Xu}
\affiliation{Department of Physics and Astronomy, University of Tennessee, Knoxville, TN 37996, USA}
\affiliation{Department of Physics, Tsinghua University, Beijing 100084, China}

\author{Nianlong Zou}
\affiliation{Department of Physics and Astronomy, University of Tennessee, Knoxville, TN 37996, USA}

\author{Nikolai Peshcherenko}%
\affiliation{%
Max Planck Institute for Chemical Physics of Solids, 01187, Dresden, Germany}

\author{Ammar Jahin}
\affiliation{Theoretical Divison, T-4, Los Alamos National Laboratory, Los Alamos, New Mexico 87545, USA}

\author{Tingxin Li}
\affiliation{School of Physics and Astronomy, Shanghai Jiao Tong
University, Shanghai, China}

\author{Shi-Zeng Lin}
\affiliation{Theoretical Divison, T-4, Los Alamos National Laboratory, Los Alamos, New Mexico 87545, USA}
\affiliation{Center for Integrated Nanotechnologies (CINT), Los Alamos National Laboratory, Los Alamos, New Mexico 87545, USA}

\author{Yang Zhang}
\email{yangzhang@utk.edu}
\affiliation{Department of Physics and Astronomy, University of Tennessee, Knoxville, TN 37996, USA}
\affiliation{Min H. Kao Department of Electrical Engineering and Computer Science, University of Tennessee, Knoxville, Tennessee 37996, USA}

\begin{abstract}

Superconductivity has been observed in twisted MoTe$_2$ within the anomalous Hall metal parent state~\cite{Xu2025SC}. Key signatures—including a fully spin/valley polarized normal state, anomalous Hall resistivity hysteresis, superconducting phase adjacent to the fractional Chern insulating state and a narrow superconducting dome at zero gating field—collectively indicate chiral superconductivity driven by intravalley pairing of electrons. Within the Kohn-Luttinger mechanism, we compute the superconducting phase diagram via random phase approximation, incorporating Coulomb repulsion in a realistic continuum model. Our results identify a dominant intravalley pairing with a narrow superconducting dome of $p+ip$ type at zero gate field. This chiral phase contrasts sharply with the much weaker time-reversal-symmetric intervalley pairing at finite gating field. Our work highlights the role of band topology in achieving robust topological superconductivity, and supports the chiral and topological nature of the superconductivity observed in twisted MoTe$_2$.
\end{abstract}

\maketitle


Transition metal dichalcogenide (TMD)-based moir\'e systems have become a pivotal platform for investigating emergent quantum phases~\cite{kennes2021moire,mak2022semiconductor}. The combination of large effective masses in TMD valence bands and the robustness of narrow moir\'e bands against twist angle disorder enables these systems to host diverse strongly correlated states. These include interaction-driven insulating phases (Mott and charge-transfer insulators)\cite{PhysRevLett.121.026402,PhysRevB.102.201115,regan2020mott,tang2019wse2,wang2020correlated,ghiotto2021quantum,li2021continuous,https://doi.org/10.48550/arxiv.2202.02055}, generalized Wigner crystal~\cite{regan2020mott,xu2020correlated,zhou2021bilayer,jin2021stripe,li2021imaging,huang2021correlated,padhi2021generalized,matty2022melting}, and quantum anomalous Hall (QAH) effects~\cite{li2021quantum}. Notably, twisted MoTe$_2$ (t-MoTe$_2$) has recently unveiled unprecedented correlated topology, with transport measurements demonstrating both integer and odd-denominator fractional QAH (FQAH) states~\cite{park2023observation,xu2023observation}, along with potential fractional quantum spin Hall signatures at $\nu= 3$~\cite{kang2024evidence}. These correlated states persist across relatively large twist angles $\theta\sim 2.5^{\circ}-3.9^{\circ}$~\cite{zeng2023thermodynamic,xu2025interplay,park2025ferromagnetism}, with complementary evidence from optical~\cite{cai2023signatures} and compressibility~\cite{zeng2023thermodynamic} measurements in the first moir\'e valence band. By contrast, twisted WSe$_2$ (t-WSe$_2$)~\cite{wang2020correlated,foutty2024mapping,knuppel2025correlated,devakul2021magic,li2021spontaneous} – an electronically analogous yet weakly correlated system – exhibits superconductivity near van Hove singularities (vHS)~\cite{guo2025superconductivity,xia2025superconductivity}, particularly in 5°-twist devices within a time-reversal-symmetric parent state.

Most recently, a surprising superconducting phase ~\cite{Xu2025SC} at $\theta\sim 3.8^{\circ}$ t-MoTe$_2$ was discovered within the anomalous Hall metal region. While the origin of superconductivity in moir\'e systems remains debated since its initial observation in twisted bilayer graphene, prevailing theories assume pairing between opposite valleys, mediated by mechanisms such as electron-phonon coupling, spin/charge fluctuations near correlated insulating phases (e.g., Mott or charge-density-wave states), or enhanced interactions at van Hove singularities. Kohn-Luttinger type of superconductivity in Fermi liquids with purely long-range repulsive interactions~\cite{Kohn1965} has also been discussed in various systems \cite{cea2021coulomb,schrade2024nematic,zhu2025superconductivity,qin2024kohn,guerci2024topological,chou2024intravalley,jahin2024enhanced,shavit2024quantum,geier2024chiral}, where effective Cooper pairing emerges from overscreening of the Coulomb potential. Recent experiments in spin-valley-polarized rhombohedral graphene~\cite{han2024signatures} and t-MoTe$_2$~\cite{Xu2025SC} challenge conventional wisdom by demonstrating superconductivity in the absence of time-reversal-symmetric parent states, prompting theoretical proposals for intravalley pairing~\cite {chou2024intravalley,jahin2024enhanced,geier2024chiral}. Notably, nontrivial band topology has been shown to amplify intravalley pairing~\cite{jahin2024enhanced,shavit2024quantum}, positioning t-MoTe$_2$ and rhombohedral graphene—systems hosting fractional Chern insulator (FCI) at commensurate fillings~\cite{cai2023signatures,zeng2023thermodynamic,park2023observation,xu2023observation,Xu2025SC}—as a prime candidate for quantum geometrically enhanced superconductivity.

In this work, we compute the Kohn-Luttinger superconducting phase diagram of t-MoTe$_2$ using the DFT-derived low-energy continuum Hamiltonian and long-range repulsive Coulomb interaction. Our analysis reveals two distinct phases: (1) A narrow $p+ip$ chiral superconducting dome at zero gating field, driven by intravalley pairing in a spin-polarized valley with vanishing moir\'e momentum, consistent with the experimental unconventional superconductivity in 3.8$^{\circ}$ t-MoTe$_2$. Remarkably, the critical temperature $T_c$ in this channel reaches Kelvin-scale values without being pinned to van Hove singularity. Similar features are also corroborated in an effective Skyrmion lattice model. (2) A wide but much weaker intervalley pairing phase with $T_c$ $\sim 10–100$ mK transitioning from $d$ to $f$-wave under gating field, similar to prior expectations for repulsion-driven Kohn-Luttinger superconductivity~\cite{schrade2024nematic,guerci2024topological,qin2024kohn}. This stark contrast 
not only supports the chiral nature of experimentally observed superconductivity in t-MoTe$_2$, but also underscores the interplay of quantum topology and Coulomb screening for stabilizing spin-triplet pairing within topologically nontrivial bands.

\begin{figure}
\includegraphics[width=\columnwidth]{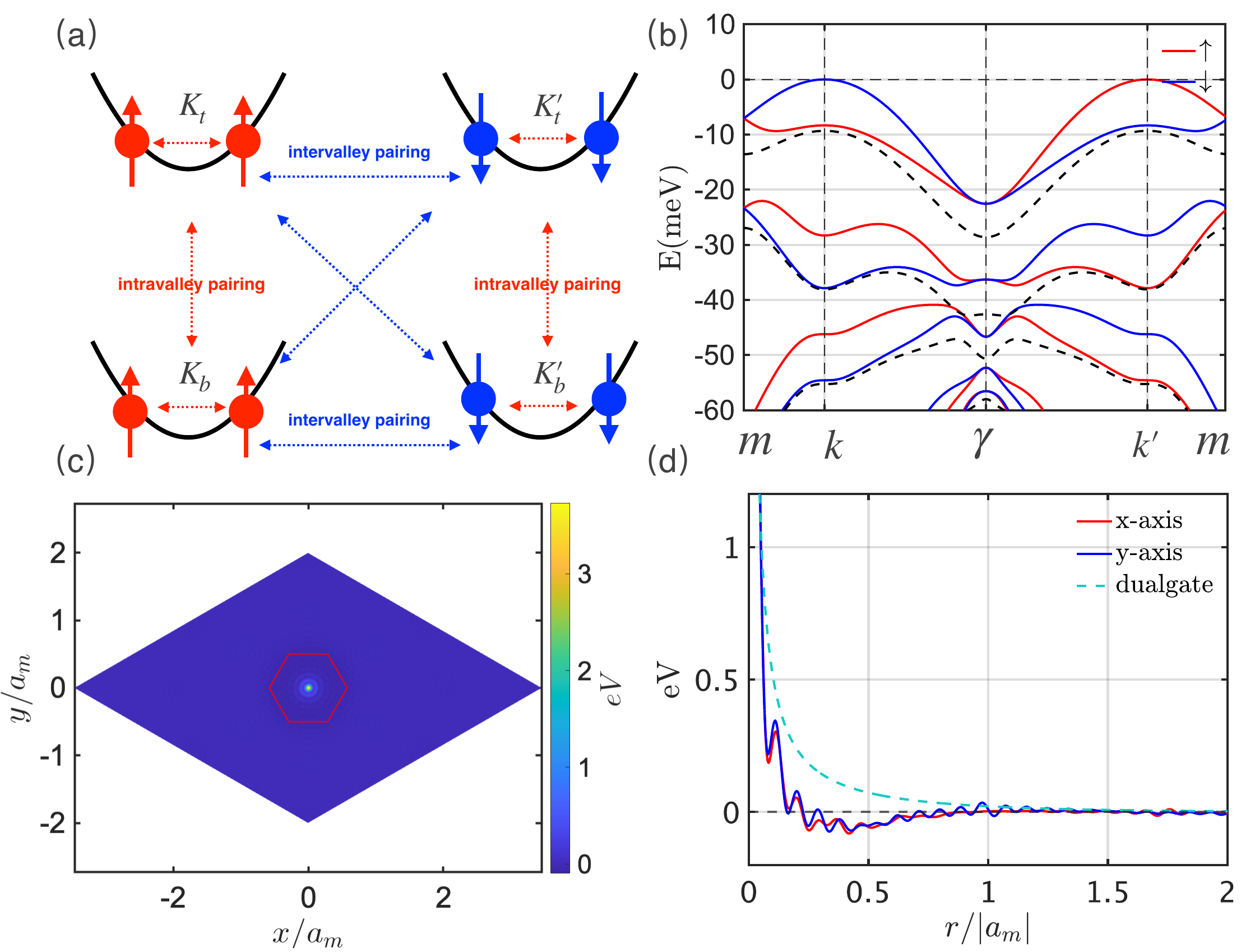}
\caption{(a) A schematic illustration of intervalley and intravalley pairing. Here, \(K_t\) and \(K_b\) denote the valleys of the top and bottom layers, respectively. The blue arrow indicates intervalley pairing, while the red arrow shows intravalley pairing. Here the gating field suppresses the intravalley pairing. (b) Band structure of t-MoTe\(_2\) at \(\theta = 3.89^\circ\), calculated using the continuum model. The black dashed line corresponds to zero displacement field, while the red (blue) lines represent spin-up (spin-down) bands under a displacement field \(\Delta = 10\) meV. (c) Real-space distribution of the RPA-screened Coulomb interaction, calculated within the single-band approximation at the van Hove filling and dielectric constant \(\epsilon = 5\). (d) The real space distribution of the interaction measured relative to the MM stacking region: blue (red) curves show the RPA-screened potential along the \(x\) (\(y\)) direction, and the dashed line denotes the unscreened dual-gate Coulomb potential.
}\label{fig1}
\end{figure}

\textbf{Continuum model electronic structure.} 
Our work begins with the continuum model \cite{wu_topological_2019} derived from DFT calculations \cite{xu2024maximally,Mao2024,osti_2356863,zhang2024polarization}. In our implementation, we have included higher-order harmonic terms in both inter-layer and intra-layer couplings, explicitly extending up to second-order harmonics. Due to the significant momentum separation between the two valleys, intervalley coupling can be safely neglected, allowing us to exclusively focus on the K valley. Furthermore, given the pronounced Ising spin-orbit coupling in MoTe$_2$, our analysis is further simplified by considering only a single spin component. Consequently, the continuum model for the K valley adopted in our calculations is given as follows:
\begin{equation}
\hat{H_s}=	\begin{bmatrix}
		-\frac{\boldsymbol{(\hat k-K_{t})^2}}{2m^*}+\Delta_{+}(\boldsymbol{r})& \Delta_{T}(\boldsymbol{r})\\
		\Delta^{\dagger}_{T}(\boldsymbol{r}) & -\frac{\boldsymbol{(\hat k-K_{b})^2}}{2m^*}+\Delta_{-}(\boldsymbol{r})
	\end{bmatrix}	
\end{equation}
with:
\begin{equation}
	\begin{aligned}
		\Delta_{\pm}(\boldsymbol{r})&=2V_1\sum_{i=1,3,5}\cos(\boldsymbol{g^{1}_i\cdot r}\pm \phi_1)+2V_2\sum_{i=1,3,5}\cos(\boldsymbol{g^{2}_i\cdot r})\\
		\Delta_{T}&=w_1\sum_{i=1,2,3}e^{-i\boldsymbol{q^1_i\cdot r}}+w_2\sum_{i=1,2,3}e^{-i\boldsymbol{q^2_i\cdot r}}\\
	\end{aligned}
\end{equation}
where $\boldsymbol{\hat{k}}$ denotes the momentum operator; $\boldsymbol{K_{t}}$ ($\boldsymbol{K_{b}}$) represents the high-symmetry momentum $\boldsymbol{K}$ of the top (bottom) layer; $\Delta_{+}(\boldsymbol{r})$ and $\Delta_{-}(\boldsymbol{r})$ describe the moir\'e potentials for the top layer and the bottom layer, respectively; $\Delta_{T}(\boldsymbol{r})$ corresponds to the interlayer tunneling potential; and $\boldsymbol{G_i}$ is a moir\'e reciprocal lattice vector. Furthermore, $\boldsymbol{g_i^1}$ and $\boldsymbol{g_i^2}$ represent momentum differences between the nearest and second-nearest plane-wave bases within the same layer, while $\boldsymbol{q_i^1}$ and $\boldsymbol{q_i^2}$ denote momentum differences between nearest and second-nearest plane-wave bases across different layers. The continuum parameters employed in our calculation are from density-dependent vdW corrections~\cite{Mao2024}: $m^*=0.62m_e$, $V_1=10.3$ meV, $V_2=2.9$ meV, $w_1=-7.8$ meV, $w_2=6.9$ meV, and $\theta = 3.89^\circ$.

\textbf{Random phase approximation.} 
After specifying the electronic structure of the continuum model, we proceed to investigate Kohn-Luttinger-type superconductivity in the context of hole doping. The essence of the Kohn-Luttinger mechanism lies in the emergence of effective attractive interactions arising from many-body quantum effects. These interactions can be captured within the framework of the random phase approximation (RPA) \cite{Gell-Mann1957Apr, Bohm1957Jul}, which involves the infinite summation of bubble diagrams. For the continuum models, the RPA-screened Coulomb interaction \( V^{\text{RPA}} \) is expressed as:
\begin{equation}
    [V^{\text{RPA}} (\mathbf{q})]^{- 1}_{\bm{G, G'}} = V^{-1}_{0}
   (\mathbf{q+G})\delta_{\bm{G, G'}} - \Pi_{\bm{G, G'}} (\mathbf{q})
\end{equation}
where the unscreening Coulomb interaction \( V^0(\mathbf{q}) \) is taken as a dual-gate screened Coulomb potential defined by $V_0(\bm{q})= \frac{e^2\tanh(|\mathbf{q}|d_s)}{2\epsilon_0\epsilon_r|\mathbf{q}|}$, where the screening length $d_s$ is set to be 10 nm and \( \Pi_{\bm{G, G'}} (\mathbf{q}) \) represents the Lindhard dielectric function. The Lindhard function for a non-interacting continuum model takes the following form:
\begin{equation}
    \Pi_{\bm{G, G'}} (\mathbf{q}) = \frac{1}{A} \sum_{k,\sigma} \frac{f^{m,
   \sigma}_{\mathbf{k}} - f^{n, \sigma}_{\mathbf{k + q}}}{\epsilon^{m,
   \sigma}_{\mathbf{k}} - \epsilon^{n, \sigma}_{\mathbf{k} +
   \mathbf{q}}} \Lambda^{m, n, \sigma}_{\mathbf{k}, \mathbf{k + q + G}}
   \Lambda^{n, m, \sigma}_{\mathbf{k + q + G}', \mathbf{k}}
\end{equation}
Here, \(\epsilon_{\bm{k}}^{m\sigma} = E_{m\bm{k}\sigma} - \mu\), with \(\mu\) denoting the chemical potential and \(E_{m\bm{k}\sigma}\) denotes the eigenenergy obtained from the continuum model; \(\sigma\) labels the spin. \( A \) denotes the system area, \( f^{m,\sigma}_\mathbf{k} \) is the Fermi–Dirac distribution function corresponding to the \( m \)-th energy band at momentum \( \mathbf{k} \), and \( \Lambda^{m, n, \sigma}_{\mathbf{k}, \mathbf{k + q + G}} = \langle m, \mathbf{k}, \sigma | n, \mathbf{k + q + G}, \sigma \rangle \) is the form factor, which encodes the influence of the band geometry on the effective many-body interactions.

Since our focus is on fillings within the first moir\'e band, which is well separated from other bands in Fig.~1(b), we adopt a single-band approximation in the main text and omit the band index accordingly (see Supplementary Material for results including additional bands). Within this approximation, the RPA-renormalized Coulomb interaction in real space is obtained, as shown in Fig.~\ref{fig1}(c). Remarkably, this approach yields an attractive component, depicted in Fig.~\ref{fig1}(d).

Using this RPA-screened Coulomb interaction and considering only zero center-of-mass momentum within the moiré Brillouin zone, the interaction projected onto the first moiré band can be written as:
\begin{equation}	
V^{\bm{\sigma,\sigma^\prime}}_{\bm{k,k'}}=\sum_{\bm{GG^\prime}}	V^{\text{RPA}}_{\bm{G,G^\prime}}(\bm{k-k'}) \Lambda^{\sigma}_{\bm{k,k'-G}}\Lambda^{\sigma^\prime}_{\bm{-k,-k'+G^\prime}}
\end{equation}
Within the generalized BCS framework, the superconductivity critical temperature $T_c$ can be obtained by solving the following linearized gap equation.
\begin{equation}
	\phi^{\sigma,\sigma^\prime}_{\bm{k}} = \sum_{\bm{k^\prime}} \mathcal{K}^{\sigma,\sigma^\prime}_{\bm{k,k^\prime}}(T) \phi^{\sigma,\sigma^\prime}_{\bm{k^\prime}}
\end{equation}
where $\mathcal{K}^{\sigma,\sigma^\prime}_{\bm{k,k^\prime}}(T)$ is the pairing kernel determined by $V^{\bm{\sigma,\sigma^\prime}}_{\bm{k,k'}}$ and temperature dependent fermi distribution function (see Supplementary Material for explicit form). $T_c$ corresponds to the solution of the equation $\lambda(T) = 1$, where $\lambda(T)$ denotes the largest eigenvalue of the kernel matrix $\mathcal{K}(T)$.
\begin{figure}
\includegraphics[width=\columnwidth]{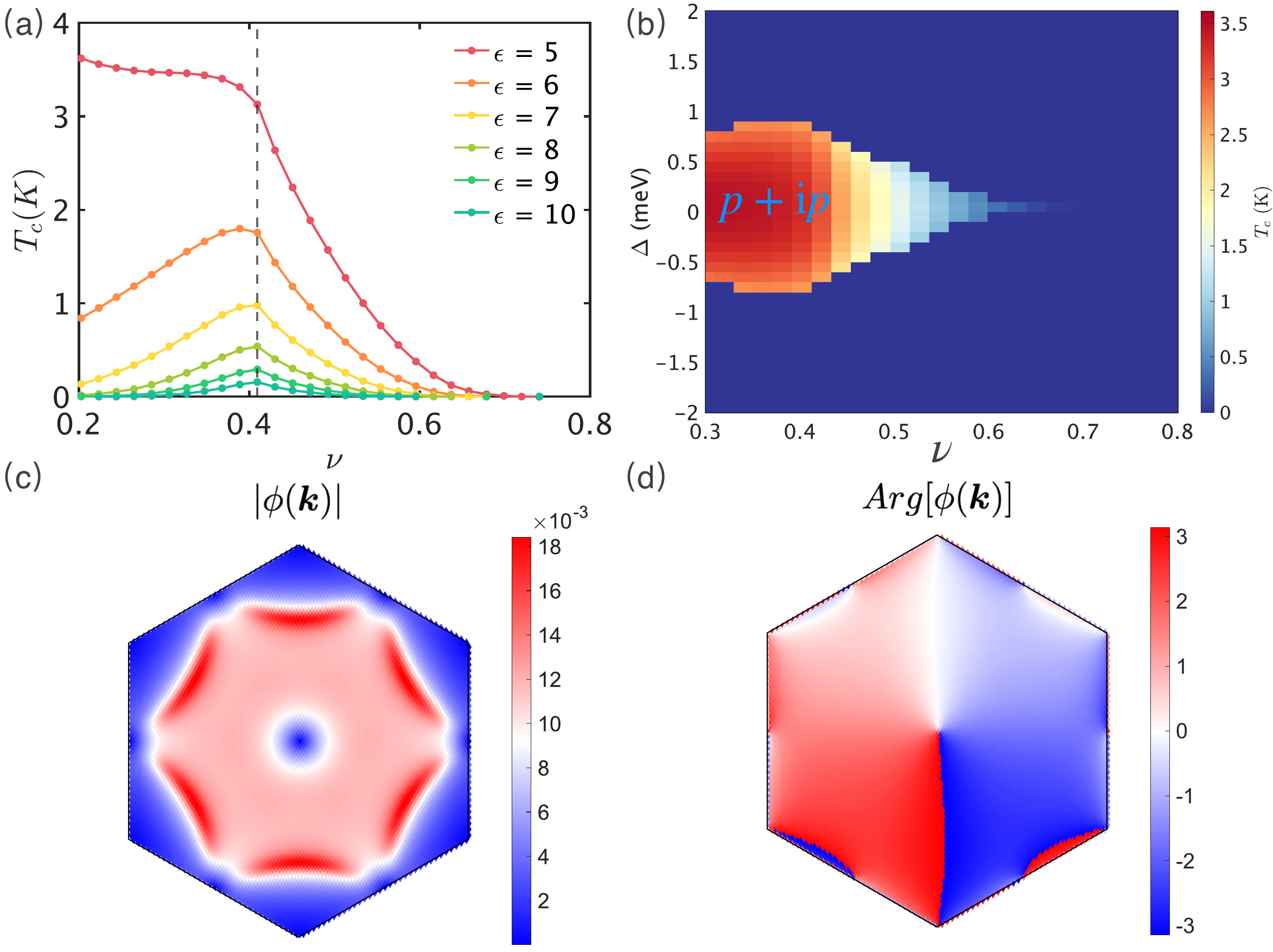}
\caption{Intravalley pairing: (a) The dependence of the critical temperature on the filling factor for various dielectric constants at zero gating field. The dashed line represents the position of van Hove singularity. (b) The critical temperature as a function of both the filling factor and the layer onsite energy difference induced by the displacement field. (c) and (d) show the magnitude and phase of the superconducting order parameter at a dielectric constant $\epsilon = 5$ and filling factor corresponding to the van Hove singularity, clearly revealing a distinct $p+ip$ pairing symmetry. }\label{fig2}
\end{figure}

\textbf{Intravalley pairing.} 
When the first moir\'e band is fractionally filled, time reversal symmetry (TRS) is typically broken through spontaneous valley polarization. This symmetry breaking leads to the experimental observation of a fractional Chern insulator with ferromagnetic order. Motivated by these findings, our theoretical framework begins with spin/valley-polarized states and examines intravalley pairing mechanisms. To streamline our analysis, the Lindhard function is computed exclusively for a single spin component. The central results are illustrated in Fig.~\ref{fig2}. First, as shown in Fig.~\ref{fig2}(a), the peak of the critical temperature does not always coincide with the van Hove singularity ($\nu_{\mathrm{vHS}} \approx 0.4$). When the dielectric constant $\epsilon$ exceeds 5, the peak indeed appears at $\nu_{\mathrm{vHS}}$. However, gradually lowering $\epsilon$ significantly enhances the critical temperature at lower filling factors. This behavior highlights the pivotal role of the form factor, which was not included in previous calculations. Secondly, as illustrated by the phase diagram in Fig.~\ref{fig2}(b), a narrow superconducting dome appears near zero gate field. Across the entire parameter range depicted in Fig.~\ref{fig2}(b), the system exhibits a $p+ip$ pairing symmetry, as illustrated by Fig.~\ref{fig2}(c) and \ref{fig2}(d).

When the onsite layer energy difference exceeds 1 meV, superconductivity completely disappears, in agreement with experimental observations~\cite{Xu2025SC}. A simple symmetry argument clarifies this behavior: Due to pseudo-inversion symmetry in continuum model, states \(| \mathbf{k} + K,\uparrow\rangle  \) and \( |-\mathbf{k} + K ,\uparrow \rangle\) within a given \(K\) valley are degenerate, enabling intravalley pairing. In t-MoTe\(_2\), gating fields lift this degeneracy, contributing to the suppression of intravalley pairing, as shown in Fig.~2(b). 
This is distinct from rhombohedral graphene~\cite{han2024signatures} where states  \(| \mathbf{k} + K,\uparrow\rangle  \) and \( |-\mathbf{k} + K ,\uparrow \rangle\) are not at the same energy due to the trigonal warping. 

\begin{figure}
\includegraphics[width=\columnwidth]{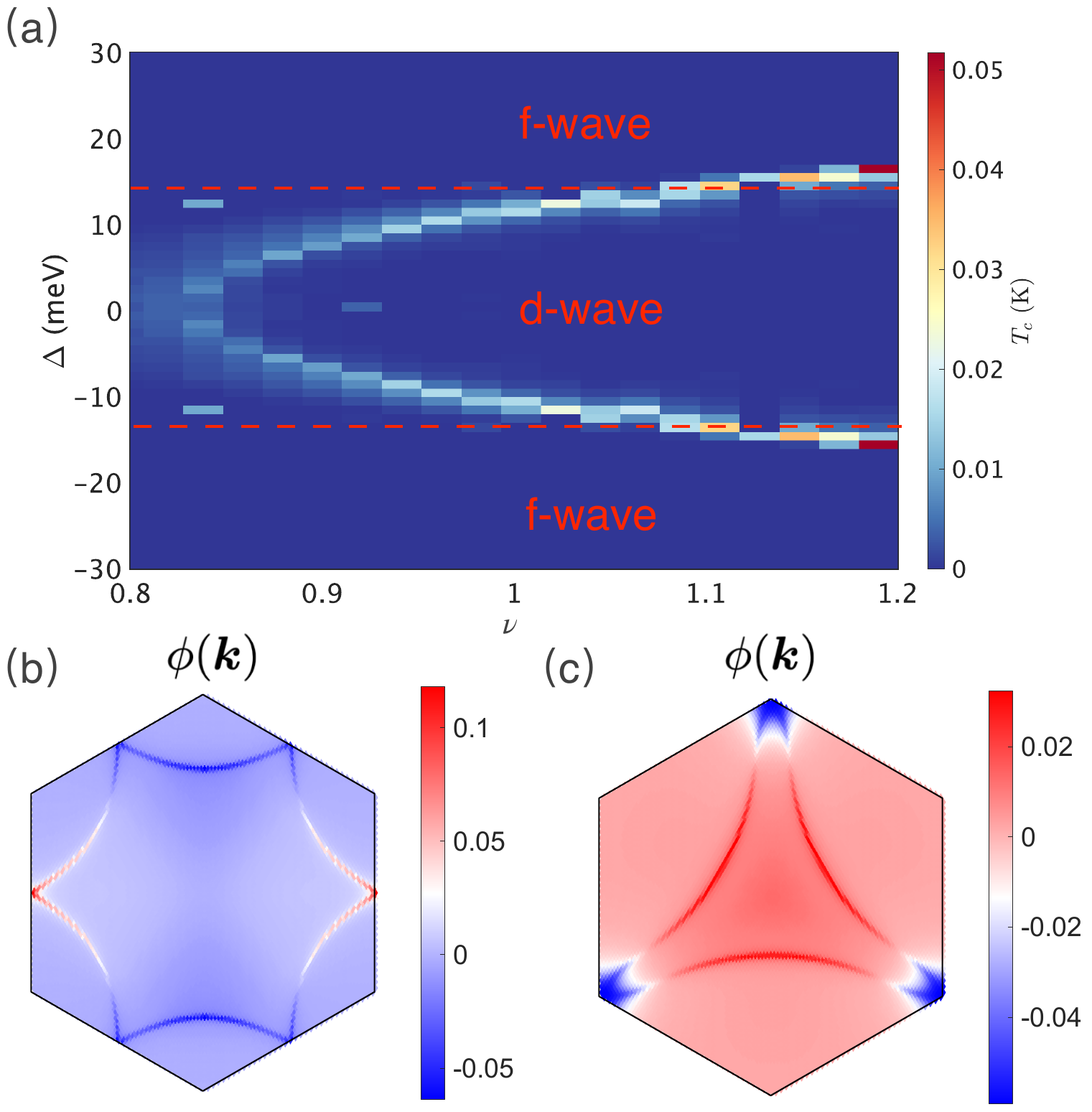}
\caption{Intervalley pairing: (a) The critical temperature as a function of both the filling factor and the layer onsite energy difference induced by the displacement field. (b) The order parameter distribution at $\Delta = 0 $ meV and $\epsilon = 5$, with the filling factor chosen at the van Hove singularity. (c) The distribution of order parameter at $\Delta = 20 $ meV and $\epsilon = 5$ with the filling factor chosen at the van Hove singularity.}\label{fig3}
\end{figure}


\textbf{Weak intervalley pairing.} 
Unlike the intravalley pairing, time-reversal symmetry is enforced in this section for the intervalley pairing simulation. Because of the similar continuum model, the intervalley pairing in t-MoTe$_2$ closely resembles that in t-WSe$_2$, as reported in recent work~\cite{schrade2024nematic,guerci2024topological,qin2024kohn}. Both systems display exceptionally low critical temperatures—on the order of 10 mK, as shown in Fig.~\ref{fig3}(a), with the superconducting dome persistently anchored at the van Hove singularity. The time-reversal symmetry ensures that the states \( |\mathbf{k} ,\uparrow\rangle\) and \(| -\mathbf{k},\downarrow\rangle \) remain degenerate in energy, a feature robust against gating fields as demonstrated in Fig.~\ref{fig1}(b). The superconductivity can survive even with the large gating field as a result. Notably, the intervalley pairing symmetry exhibits distinct characteristics compared to intravalley pairing. At weak gating fields, the system adopts a $d$-wave symmetry, as evidenced by the momentum-space structure of the order parameter in Fig.~\ref{fig3}(b). As the gating field increases, a symmetry crossover occurs, driving the system into a $f$-wave pairing symmetry in Fig.~\ref{fig3}(c). This transition reflects the interplay between field-induced band deformation and the competing energy scales of pairing interactions, consistent with the theoretical calculation in t-WSe$_2$~\cite{schrade2024nematic,guerci2024topological,qin2024kohn}.

\begin{figure}
\includegraphics[width=\columnwidth]{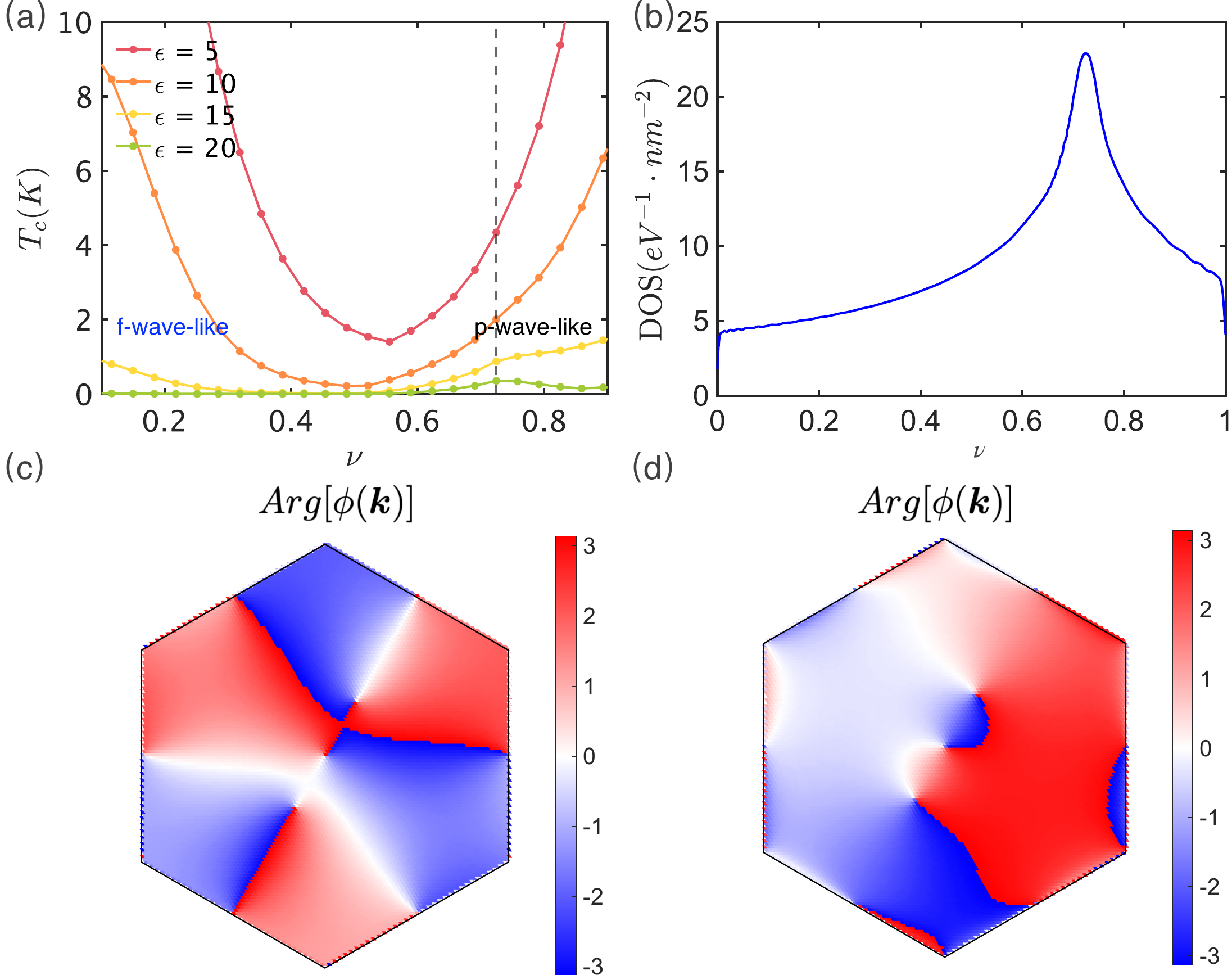}
\caption{Skyrmion model: (a) The dependence of the critical temperature on the filling factor for various dielectric constants. The dashed line represents the position of van Hove singularity. (b) The density of states of the 1st band of Skyrmion lattice model. (c) The phase distribution of the order parameter at $\nu=0.3$ and $\epsilon=5$. (d) The phase distribution of the order parameter at $\nu=0.7$ and $\epsilon = 5$. 
}\label{fig5}

\end{figure}

\textbf{Skyrmion lattice model.} 
The band structure of t-MoTe$_2$ stands out for its Landau-level-like quantum geometry \cite{PhysRevB.109.245131,PhysRevLett.132.096602,Reddy2024NonAbelianFI, xu2025multiple,Ahn2024NonAbelianFQ,wang2025higher} — an intrinsically nontrivial property that underpins topological phases such as the QAH and FQAH. Within our framework, this quantum geometry shapes both the Lindhard function and the pairing potential. Notably, in the intravalley scenario, the pairing potential is affected not just by the magnitude of the form factor but also by its phase, which encapsulates the Landau-level-like character. Inspired by these findings, we turn to an effective Landau-level-like model---namely, the Skyrmion lattice model: 
\begin{equation}
	\hat{H}=\frac{\hat{p}^2}{2m}+J\bm{ \sigma\cdot S(r)}
\end{equation}
with:
\begin{equation}
\begin{aligned}
&	\bm{S(\bm{r})}=\frac{\bm{N(r)}}{N(\bm{r})}\\	
&	\bm{N(r)}=\frac{1}{\sqrt{2}}\sum_{j=1}^6 e^{i\bm{q_j\cdot r}}\hat{e}_j+N_0\hat{z}\\
&    \hat{e}_j=(i\alpha \sin \theta_j,-i\alpha\cos\theta_j,-1)/\sqrt{2}\\
& \bm{q_j}=\frac{4\pi}{\sqrt{3}a_m}(\cos\theta_j,\sin\theta_j)
\end{aligned}	
\end{equation}
For all of the calculations below, we set the following parameters: $m=0.6m_e$, $\alpha=1$, $N_0=0.28$, $J=0.5 eV$, and $a_m=50\AA$. We focus on its lowest band, considering pairing between \(\bm{k}\) and \(-\bm{k}\), analogous to the intravalley case discussed above. In t-MoTe\(_2\), we already observe that the peak in \(T_c\) shifts away from the van Hove singularity, as shown in Fig.~\ref{fig2}(a). Remarkably, in the Skyrmion lattice model, the peak of \(T_c\) is no longer tied to the location of the peak of density of states.

Furthermore, at both low filling factors near 0 and high filling factors near 1, \(T_c\) reaches high values when the dielectric constant is small, mirroring the behavior seen in t-MoTe\(_2\). This observation suggests that the system’s nontrivial band topology can surpass the constraints of a purely density-of-states-based approach—which would otherwise predict a low \(T_c\). Conversely, for sufficiently large dielectric constants (\(>15\)), the peak shifts back to the vHS. The distribution of the order parameter also warrants attention. For each curve in Fig.~\ref{fig5}(a), the pairing symmetry transitions from an \(f\)-wave-like order parameter at lower filling factors, resembling Fig.~\ref{fig5}(c), to a \(p\)-wave-like pairing symmetry at higher filling factors as we show in Fig.~\ref{fig5}(d).

While the random phase approximation employed in this work provides critical insights, its neglect of exchange corrections may quantitatively underestimate the critical temperature $T_c$ for intervalley pairing by suppressing spin/valley fluctuations. However, such limitations have minimal impact on the spin-polarized intravalley pairing mechanism central to this work. The current RPA framework also struggles to account for finer experimental details, such as the observed superconductivity onset at filling $\nu=0.7$. This discrepancy likely stems from the single-step RPA approach, which omits interaction-driven band renormalization effects. Such renormalizations could modify the doping-dependent phase boundaries in Fig.~2(b), shifting the precise location of the superconducting dome. Despite these quantitative challenges, our theory robustly captures the qualitative competition between the anomalous Hall metal and superconducting phases, offering a foundational model for future studies incorporating dynamical screening and self-energy corrections.

\textbf{Relation to ideal Chern band.}
The form factor $\Lambda$, encoding band geometry and topology, critically governs intravalley pairing. For intravalley pairing, the contribution of $\Lambda^{\sigma}_{\bm{k_1,k_4-G}}\Lambda^{\sigma^\prime}_{\bm{-k_1,-k_4+G^\prime}}$ is complex regardless of gauge choice. This lifts the degeneracy between $l$ and $-l$ angular momentum pairing channel. As shown in Ref. \cite{jahin2024enhanced}, ideal topological bands (modeled by lowest Landau level form factors) host robust chiral superconductivity, with $T_c$ dictated by Berry flux through the Fermi surface. Intravalley pairing dominates exponentially over intervalley processes, where contribution of $\Lambda^{\sigma}_{\bm{k_1,k_4-G}}\Lambda^{\sigma^\prime}_{\bm{-k_1,-k_4+G^\prime}}$ is positive real and insensitive to Berry curvature—highlighting its pivotal role in pairing enhancement.

The interplay between van Hove singularities and band topology governs the superconducting dome in t-MoTe\(_2\): while vHS enhances intervalley pairing $T_c$ at their characteristic fillings, the form factor shifts the optimal intravalley $T_c$ to distinct fillings due to competing band topology and density-of-states effects as shown in Fig.~2(a) and Fig.~4(a). Weak Coulomb interactions (large $\epsilon$) favor vHS-dominated pairing, but stronger interactions drive a crossover where $T_c$ peaks at fillings balancing vHS and Berry flux~\cite{jahin2024enhanced}. 



\textbf{Discussion.} The observation of unconventional superconductivity in t-MoTe$_2$~\cite{Xu2025SC} emerging within a fully spin/valley-polarized anomalous Hall metal regime marks a significant departure from conventional pairing mechanisms. Key experimental signatures, such as full valley polarization with anomalous hysteresis loops of resistance versus magnetic field in the normal state, 
and superconducting phase adjacent to FQAH, collectively, point to a chiral superconducting phase stabilized by intravalley pairing. The fully spin/valley-polarized system studied here suppresses spin/valley fluctuations, which makes the spin or valley fluctuations less likely to mediate superconductivity. 
Our analysis assumes a Kohn-Luttinger mechanism, where charge fluctuations induce overscreening of the Coulomb interaction, generating effective attraction for intravalley pairing. The nontrivial band topology enhances this pairing, yielding phenomenology consistent with key experimental signatures: 1) dominant intravlley pairing versus intervalley pairing, 2) superconducting dome with peak of $T_c$ not pinned to filling associated with the vHS, 3) suppression of superconductivity by vertical electric field, 4) $T_c$ of the order of $1$ K. The interplay of equal-spin pairing and the Chern band’s topology stabilizes a chiral $p+ip$-wave superconducting state. One immediate consequence is the appearance of Majorana fermions localized at the vortex core and chiral Majorana mode circulating at the sample edge~\cite{kallin2016chiral,ivanov2001non,venderbos2016odd}—key ingredients for fault-tolerant topological quantum computing.

The TRS breaking order parameters of chiral superconductivity can be directly detected by optical sensing, such as magnetic circular dichroism and Kerr rotation. Additionally, scanning superconducting quantum interference device (SQUID) microscopy could map spontaneous edge or domain-boundary magnetic fluxes—a hallmark of chiral superconductivity. Unlike rhombohedral graphene, the superconducting phase in t-MoTe2 appears at zero electric field, which can facilitate the STM measurement to detect the Majorana fermions inside the vortex core.

The recent discovery of zero-field integer and fractional quantum Hall effects in t-MoTe$_2$ \cite{park2023observation,xu2023observation} and pentalayer graphene \cite{lu2024fractional} establishes these systems as ideal platforms for exploring Chern bands and charge fractionalization without extreme magnetic fields. Strikingly, in t-MoTe$_2$, superconducting and FCI phases coexist within the same device—a phenomenon absent in Landau-level systems. This immediately raises an important question about the competition between FCI and superconductivity and the commonality of the band geometry and topology in supporting these two states. Our continuum model calculations using realistic material parameters for t-MoTe$_2$ show $T_c\sim 1$ K, which has the same order of magnitude as the gap of FCI. Practically, this proximity opens avenues for engineering hybrid FCI-superconductor heterostructures to host non-Abelian parafermions~\cite{Alicea_Fendley_2016,Liu2025para}. 
\SZL{

}

\textbf{Acknowledgments}
We are grateful to Ning Mao, Zimeng Zeng, and Cristian Batista for helpful discussions, Wei Qin and Fengcheng Wu for the cross-check of intervalley superconducting phase diagram in t-WSe$_2$~\cite{qin2024kohn}. C. X., N. Z. and Y. Z. are supported by the start-up fund at University of Tennessee Knoxville. The work at LANL was carried out under the auspices of the U.S. DOE NNSA under contract No. 89233218CNA000001 through the LDRD Program, and was supported by the Center for Nonlinear Studies at LANL, and was performed, in part, at the Center for Integrated Nanotechnologies, an Office of Science User Facility operated for the U.S. DOE Office of Science, under user proposals $\#2018BU0010$ and $\#2018BU0083$.

\bibliography{main}

\end{document}